\documentclass[intlimits,twoside,a4paper]{article}

\usepackage{graphicx}
\usepackage[cp1251]{inputenc}

\usepackage[eqsecnum]{cmpj3}

\issue{2017}{20}{2}{23001}
\doinumber{10.5488/CMP.20.23001}

\title[Generalization of the Grad method in plasma physics]%
{Generalization of the Grad method in plasma physics}
\author[V.N. Gorev,  A.I. Sokolovsky]{V.N. Gorev,  A.I. Sokolovsky}
\address{ Oles Honchar Dnipro National University, 72 Gagarin Ave., 49010 Dnipro, Ukraine}

\date{Received December 24, 2016, in final form April 5, 2017}
\begin{document}

\maketitle

\begin{abstract}
The Grad method is generalized based on the Bogolyubov idea of the functional hypothesis for states at the end of relaxation processes in a system. The Grad problem (i.e., description of the Maxwell relaxation) for a completely ionized spatially uniform two-component electron-ion plasma is investigated using the Landau  kinetic equation. The component distribution functions and time evolution equations for parameters describing the state of a system are calculated, and corrections are obtained to the known results in a perturbation theory in a small electron-to-ion mass ratio.
\pacs 02.30.Rz, 05.20.Dd,  51.10.+y, 52.25.Dg
\end{abstract}

\section{Introduction}
This paper is devoted to a generalization of the Grad method \cite{Grad49,Silin} in the physics of plasma. Although the Grad method was proposed by Grad in 1949, that method and its modifications are still widely used in modern statistical physics, for example, in the theory of granular materials \cite{Garzo13,Garzo15}, in the investigation of relativistic hydrodynamics \cite{DerradiDeSouza}, in the investigations of the shock wave structure \cite{ShockWave}, in the physics of plasma \cite{Zhdanov,Zhdanov2}, etc. In this paper we concentrate on the 13-moment Grad approximation. This approximation describes the formation of dissipative hydrodynamic fluxes (the Maxwell relaxation), and thus it is very important and popular.

Within the framework of the 13-moment Grad approximation, the reduced description parameters (RDPs) (i.e., parameters describing the state of a system) are the component particle densities $n_a$, component velocities~$\upsilon_{an}$, component temperatures $T_a$, component traceless momentum fluxes $\pi^o_{anl}$ and component energy fluxes $q^o_{an}$ taken in the reference frame which accompanies the $a$-th component ($a=e,i$ is the component subscript). Under the widely used assumption \cite{Grad49,Silin,Zhdanov}, within the framework of the 13-moment approximation, the component distribution functions (CDFs) are as follows:
\begin{equation} \label{standardCDF}
{f_{ap}} = \left[ {1 + \frac{{{h_{nlp}}}}{{2{m_a}{n_a}T_a^2}}\pi _{anl}^o - \frac{2}{5}\frac{{{p_l}}}{{{n_a}T_a^2}}\left( {\frac{5}{2} - \frac{{{\varepsilon _{ap}}}}{{{T_a}}}} \right)q_{al}^o} \right]_{p_a \rightarrow p_a-m_a\upsilon _a}\cdot f_{ap}^L\,,
\end{equation}
where $ {h_{nlp}} \equiv {p_n}{p_l} - {{{p^2}{\delta _{nl}}} / 3}$,  ${\varepsilon _{ap}} \equiv {{{p^2}} / {2{m_a}}}
$ and
\begin{equation} \label{LEA_CDF}
f_{ap}^L = \frac{{{n_a}}}{{{{\left( {2\piup {m_a}T_a} \right)}^{{3 / 2}}}}}\exp \left[ { - \frac{{{{\left( {p - {m_a}{\upsilon _a}} \right)}^2}}}{{2{m_a}T_a}}} \right]
\end{equation}
are the local equilibrium CDFs. In what follows, we will use the term ``standard result'' for the expression~\eqref{standardCDF}.
The CDFs~\eqref{standardCDF} are the product of CDFs $f_{ap}^L$ and some combination of fluxes which can be obtained based on the truncated Hermite polynomial expansion \cite{Grad49,Silin} and the additional conditions, which are the definitions of the RDPs in terms of the CDFs. In usual hydrodynamic states, the fluxes $\pi^o_{anl}$, $q^o_{an}$ are values of the first order in the gradients of hydrodynamic variables, and~\eqref{standardCDF} shows that $f_{ap}^L$ gives a zero order contribution in gradients to the non-equilibrium CDFs $f_{ap}$. This result expresses the local equilibrium assumption, which is widely used in the literature (see, for example, the discussion of the temperature and velocity relaxation in a completely ionized plasma \cite{AlBogRuh,Ishi}, the discussion of temperature relaxation in an electron-phonon system \cite{Singh}, etc.).

However, in some works \cite{Sizhuk,BogPol,Sok1,Sok2}, a violation of the local equilibrium assumption in spatially uniform systems is discussed. In our paper \cite{GorSokCMP} it is also stressed that the CDFs $f_{ap}^L$ are not exact solutions of the kinetic equation, and corrections to them in a perturbation theory in a small parameter
\begin{equation} \label{sigma}
\sigma  = \sqrt {m_e/m_i}
\end{equation}
are obtained. Thus, we expect to obtain  corrections to the expression~\eqref{standardCDF} in the same perturbation theory.

As known \cite{Jou}, a drawback of the Grad method is the lack of a small parameter, which does not allow one to obtain the CDFs from the kinetic equation and that is why the CDFs are postulated. In the considered problem, within the framework of the standard Grad method, the CDFs are postulated in the form~\eqref{standardCDF}. In our paper \cite{GorSokCMP}, a system is investigated in the vicinity of its equilibrium state, and the deviations of the RDPs from their equilibrium values are considered to be small. This yields an additional small parameter, which allows one to calculate the non-equilibrium CDFs from the kinetic equation.

The use of such a small parameter that describes the deviation of states of a system from some classes of non-equilibrium states is the main feature of our approach to the theory of relaxation processes that can be observed in spatially uniform systems and can be taken into account in the theory of nonuniform systems as well \cite{GorSokNasMath,GorSokIJMPB,GorSokVis}. In other words, this is an approach to the investigation of the effect of kinetic modes of a system on its evolution. This important problem is widely discussed in the literature without the presence of a small parameter in the constructed theory (see, for example, \cite{Mryglod}).

The idea of the present paper is similar to that of the work \cite{GorSokCMP} devoted to the temperature and velocity relaxation in plasma, but here the Maxwell relaxation is taken into account too. The aim of the paper is to obtain the CDFs of the plasma based on the Landau kinetic equation and to obtain corrections to a standard result~\eqref{standardCDF} following the Grad theory. The time evolution equations for the RDPs are also obtained. In the present paper, the developed theory is restricted by a linear approximation but it can describe nonlinear relaxation processes (see paper \cite{GorSokVis} where a quadratic relaxation is discussed).

It should be noted that this paper is based on the Landau kinetic equation which can be obtained using the Bogolyubov reduced description method based on the functional hypothesis that states: at times which are much longer than the collision time, the many-particle distributions functions depend on time through the one-particle distribution function. Such a theory is valid if the gas under consideration is rarefied, and it adequately describes a completely ionized plasma. In the case of dense gases and liquids, other approaches should be used. For example, a reduced description of a system by the one-particle distribution function and the densities of hydrodynamic quantities is discussed in \cite{ZubMorOmelTok}. Some other peculiarities should be taken into account for quantum systems (see, for example, \cite{Zubarev,GerVob}). In fact, the main trend of the modern theory of nonequilibrium processes is an extension of the set of reduced description parameters. The 13-moment Grad problem discussed herein is concerned with this trend using the energy and momentum fluxes in hydrodynamics as additional independent variables.

The paper is organized as follows. In section~\ref{SectionBasic}, the basic equations of the theory are presented, and in section~\ref{CalculationOfCDFs}, the CDFs and time evolution equations for the RDPs are calculated within the framework of a linear relaxation theory.

\section{Basic equations of the theory} \label{SectionBasic}

The paper is based on the Landau kinetic equation which in the spatially uniform case is of the form
\begin{align} \label{Landau_KE}
{\partial _t}{f_{ap}} &= {I_{ap}}\left( f \right),  \\
{I_{ap}}\left( f \right) &= 2\piup e_a^2L\sum\limits_c {e_c^2\frac{\partial }{{\partial {p_n}}}} \int {{\rd^3}p'\left( {{f_{cp'}}\frac{{\partial {f_{ap}}}}{{\partial {p_k}}} - {f_{ap}}\frac{{\partial {f_{cp'}}}}{{\partial {{p'}_k}}}} \right){D_{nk}}} \left( {\frac{p}{{{m_a}}} - \frac{{p'}}{{{m_c}}}} \right), \nonumber
\end{align}
\begin{align}
D{{\kern 1pt} _{nk}}\left( u \right) &\equiv \frac{ u^2 \delta _{nk} - u_n u_k}{u^3}\,,\nonumber
\end{align}
where $f_{ap}$ are the CDFs, $I_{ap}$ is the Landau collision integral, $e_a$ is the component charge ($e_e=-e$, $e_i=ze$, $e$ is the elementary electric charge, $z$ is the ion charge number) and $L$ is the Coulomb logarithm.

Standard definitions of the RDPs in terms of the CDFs are \cite{Silin}
\begin{eqnarray} \label{RDP_definitions}
&& {n_a} = \int {{\rd^3}p\,{f_{ap}}}\,, \qquad {n_a}{m_a}{\upsilon _{an}} = \int {{\rd^3}p\,{p_n}{f_{ap}}}\,, \qquad
\frac{3}{2}{n_a}{T_a} + \frac{1}{2}{n_a}{m_a}\upsilon _a^2 = \int {{\rd^3}p\,{\varepsilon _{ap}}{f_{ap}}}\,, \nonumber \\
&& \pi _{anl}^o = \int {{\rd^3}p\,\frac{{{p_n}{p_l}}}{{{m_a}}}{f_{a,p + {m_a}{\upsilon _a}}}}\,, \qquad q_{an}^o \equiv \int {{\rd^3}p\,\frac{{{p_n}{\varepsilon _{ap}}}}{{{m_a}}}{f_{a,p + {m_a}{\upsilon _a}}}}.
\end{eqnarray}
The equilibrium temperature $T$ and velocity $\upsilon_n$ of the system are also introduced by standard definitions:
\begin{align} \label{eqilibrium_v_T}
{\upsilon _n}\sum\limits_a {{m_a}{n_a}}  &= \sum\limits_a {{m_a}{n_a}{\upsilon _{an}}}\,, \nonumber \\
 \frac{3}{2}T\sum\limits_a {{n_a}}  + \frac{1}{2}{\upsilon ^2}\sum\limits_a {{m_a}{n_a}}  &= \frac{3}{2}\sum\limits_a {{n_a}{T_a}}  + \frac{1}{2}\sum\limits_a {{m_a}{n_a}} \upsilon _a^2.
\end{align}
Based on~\eqref{Landau_KE}--\eqref{eqilibrium_v_T}, it can be shown that ${\partial _t}T = 0$, ${\partial _t}{\upsilon _n} = 0$, $\partial_t n_a=0$. In what follows we use the reference frame where $\upsilon_n=0$ and the electroneutrality condition $n_e=zn_i$.

Let us introduce the deviations of the electron temperature and velocity from their equilibrium values:
\begin{equation} \label{u_tau_definitions}
{u_n} = {\upsilon _{en}} - {\upsilon _n}\,, \qquad \tau  = {T_e} - T.
\end{equation}
As shown \cite{GorSokCMP},
\begin{equation} \label{va_ta_InTermsOf_u_tau}
{\upsilon _{an}} = {r_a}{u_n}\,, \qquad {T_a} = T + {s_a}\tau  + {y_a}{u^2},
\end{equation}
where
\begin{equation} \label{r_s_definitions}
{s_a} = {\delta _{ae}} - z{\delta _{ai}}\,, \qquad {y_a} =  - {{{m_e}z\big( {1 + z{\sigma ^2}} \big){\delta _{ai}}} / 3}, \qquad {r_a} = {\delta _{ae}} - z{\sigma ^2}{\delta _{ai}}.
\end{equation}
Formulae~\eqref{va_ta_InTermsOf_u_tau}, \eqref{r_s_definitions} express the component temperatures and velocities in terms of the deviations $\tau$, $u_n$ and equilibrium quantities. Thus, the following set of the reduced description parameters can be chosen:
\begin{equation} \label{useful_RDP}
\xi_\alpha:\qquad {u_n}\,,\,\tau,\pi _{enl}^o\,,\,\pi _{inl}^o\,,\,q_{en}^o\,,\,q_{in}^o.
\end{equation}
The developed theory is based on the Bogolyubov idea of the functional hypothesis (see a review in \cite{AhPel}), which can be written in the form:
\begin{equation} \label{functional_hypothesis}
{f_{ap}}\left( t \right)\xrightarrow[{t \gg {\tau _0}}]{}{f_{ap}}\big( {\xi \left( t \right)} \big),
\end{equation}
where $f_{ap}(\xi)$ is a function of the variables $\xi_\alpha$,  $\tau_0$ is some characteristic time which is much shorter than the shortest relaxation time of the RDPs. According to~\eqref{functional_hypothesis}, the Landau kinetic equation~\eqref{Landau_KE} can be rewritten as an equation for the CDFs $f_{ap}(\xi)$
\begin{equation} \label{rewritten_KE}
\sum\limits_\alpha  {\frac{{\partial {f_{ap}}\left( \xi  \right)}}{{\partial {\xi _\alpha }}}{L_\alpha }} \big( f( \xi)  \big) = {I_{ap}}\big( {f( \xi )}\big).
\end{equation}
Here, the time evolution equation for the RDPs is used in the form
\begin{equation} \label{time_equations}
\partial_t\xi_\alpha(t)= L_\alpha \Big(f\big(\xi(t)\big)\Big),
\end{equation}
where the function $L_\alpha(f)$ is given by the definition of the RDPs and the kinetic equation.

In the present paper, we consider the system to be in the vicinity of its equilibrium state. All the RDPs $\xi_\alpha$ vanish in the equilibrium state and we consider them to be small and estimate them by one small parameter $\mu$ for simplicity. This parameter is introduced based on the dimensional estimates
\begin{equation} \label{parameter_mu}
\tau  \sim \mu T, \qquad {u_n} \sim \mu \,(T/m_e)^{1/2}, \qquad \pi _{anl}^o \sim \mu \,nT, \qquad q_{an}^o \sim \mu \,nT (T/m_e)^{1/2}.
\end{equation}
The presence of this small parameter allows us to calculate the CDFs $f_{ap}(\xi)$ and the right-hand side $L_a\big(f(\xi)\big)$ of the time evolution equations for the RDPs from the kinetic equation~\eqref{rewritten_KE} in a perturbation theory in the small parameter $\mu$. In this investigation, the definitions of the RDPs in terms of CDFs $f_{ap}(\xi)$ (additional conditions to equation~\eqref{rewritten_KE}) must be used. In such a way, in the present paper, the Maxwell relaxation theory is built based on the Bogolyubov reduced description method.

Obviously, in the leading approximation, we obtain the Maxwell CDFs, and the RDPs are constant ones:
\begin{equation}
f_{ap}^{(0)} = w_{ap}\,, \qquad {w_{ap}} = \frac{n_a}{\left( {2\piup {m_a}T} \right)^{3/2}}\exp \left( { - \beta {\varepsilon _{ap}}} \right), \qquad L_{\alpha}^{(0)} = 0
\end{equation}
($\beta  \equiv {T^{ - 1}}$). Here and in what follows, the subscript in parentheses denotes the order in $\mu$:
\begin{equation}
{f_{ap}} = {w_{ap}} + f_{ap}^{\,(1)} + O\big( {{\mu ^2}} \big), \qquad   {L_\alpha } = L_\alpha ^{(1)} + O\big( {{\mu ^2}} \big).
\end{equation}
{\it Although in this paper we restrict ourselves to a linear relaxation theory, i.e., to a theory linear in the small parameter $\mu$, our method gives one an opportunity to obtain $f_{ap}(\xi)$ and $L_\alpha(f)$ in higher orders in~$\mu$}~\cite{GorSokVis}.

\section{Calculation of the component distribution functions and time evolution equations for reduced description parameters} \label{CalculationOfCDFs}

This section is devoted to the calculation of the CDFs within the framework of a linear relaxation theory. In the first order in $\mu$, due to the rotational invariance, the CDFs are of the form
\begin{equation} \label{f1_structure}
f_{ap}^{\,(1)} = {w_{ap}}\left( {A_{ap}^\tau \tau  + A_{ap}^u{p_nu_n} + \sum\limits_b {A_{ap}^{{\pi _b}}h_{nlp}\pi _{bnl}^o}  + \sum\limits_b {A_{ap}^{{q_b}}p_nq_{bn}^o} } \right),
\end{equation}
where $A_{ap}^\tau$, $A_{ap}^u$, $A_{ap}^{\pi_b}$ and $A_{ap}^{q_b}$ are some unknown functions which should be calculated. Based on~\eqref{rewritten_KE} and~\eqref{f1_structure} it can be shown that the RDP time evolution equations~\eqref{time_equations} in the first order in $\mu$ are of the form
\begin{eqnarray} \label{mu1_time_equations}
&& {\left( {{\partial _t}\tau } \right)^{(1)}} =  - {\lambda _T}\tau,  \qquad {\left( {{\partial _t}{u_l}} \right)^{(1)}} =  - {\lambda _{uu}}{u_l} - \sum\limits_b {{\lambda _{u{q_b}}}q_{bl}^o} \, , \nonumber \\
&&{\left( {{\partial _t}\pi _{anl}^o} \right)^{(1)}} =  -2\sum\limits_b {{\lambda _{{\pi _a}{\pi _b}}}\pi _{bnl}^o}\,
, \qquad {\left( {{\partial _t}q_{al}^o} \right)^{(1)}} =  - {\lambda _{{q_a}u}}{u_l} - \sum\limits_b {{\lambda _{{q_a}{q_b}}}} q_{bl}^o\,,
\end{eqnarray}
where the relaxation constants can be written as:
\begin{align} \label{mu1_relaxation_constants}
& {\lambda _{{\pi _a}{\pi _b}}} = \frac{1}{{10{m_a}}}\sum\limits_c {{{\left\{ {{h_{nlp}},{h_{nlp}}A_{cp}^{{\pi _b}}} \right\}}_{ac}}}\,, \qquad {\lambda _{u{q_b}}} = \frac{1}{{3{m_e}{n_e}}}\sum\limits_a {{{\{ {p_n},{p_n}A_{ap}^{{q_b}}\} }_{ea}}}\, ,  \nonumber \\
& {\lambda _{uu}} = \frac{1}{{3{m_e}{n_e}}}\sum\limits_a {{{\{ {p_n},{p_n}A_{ap}^u\} }_{ea}}} \,,\qquad
 {\lambda _{{q_a}u}} = \frac{1}{{3{m_a}}}\sum\limits_b {{{\left\{ {{\varepsilon _{ap}}{p_n},{p_n}A_{bp}^u} \right\}}_{ab}}}  - \frac{5}{2}{r_a}{n_a}T{\lambda _{uu}}\,, \nonumber \\
&{\lambda _{{q_a}{q_b}}} = \frac{1}{{3{m_a}}}\sum\limits_c {{{\left\{ {{\varepsilon _{ap}}{p_n},{p_n}A_{cp}^{{q_b}}} \right\}}_{ac}}}  - \frac{5}{2}{r_a}{n_a}T{\lambda _{u{q_b}}}.
\end{align}
Here, the integral brackets $\left\{ g,h \right\}_{ab}$ and the operator of the linearized collision integral $\hat K _{ab}$ are introduced:
\begin{eqnarray} \label{bracket_definition}
&& {M_{ab}}\left( {p,p'} \right) \equiv {\left. {\left( {{{\delta {I_{ap}}} \mathord{\left/
 {\vphantom {{\delta {I_{ap}}} {\delta {f_{bp'}}}}} \right.
 \kern-\nulldelimiterspace} {\delta {f_{bp'}}}}} \right)} \right|_{{f_p} = {w_p}}}\,, \qquad {M_{ab}}\left( {p,p'} \right){w_{bp'}} \equiv  - {w_{ap}}{K_{ab}}\left( {p,p'} \right), \nonumber \\
&& {\hat K_{ab}}{h_p} \equiv \int {{\rd^3}p'{K_{ab}}\left( {p,p'} \right){h_{p'}}}\,, \qquad \left\{ {{g_p},{h_p}} \right\}_{ab} \equiv \int {{\rd^3}p\,{w_{ap}}{g_p}{{\hat K}_{ab}}{h_p}}.
\end{eqnarray}
An explicit expression for the operator of the linearized collision integral can be obtained from~\eqref{Landau_KE} and its definition~\eqref{bracket_definition}:
\begin{eqnarray} \label{K_explicit}
{\hat K_{ab}}{h_p} = 2\piup e_a^2Lw_{ap}^{ - 1}\frac{\partial }{{\partial {p_n}}}\sum\limits_c {e_c^2} \int {{\rd^3}p'} \left( {{\delta _{bc}}\frac{{\partial {h_{p'}}}}{{\partial {{p'}_l}}} - {\delta _{ab}}\frac{{\partial {h_p}}}{{\partial {p_l}}}} \right){w_{ap}}{w_{cp'}} {D_{nl}}\left( {\frac{p}{{{m_a}}} - \frac{{p'}}{{{m_c}}}} \right).
\end{eqnarray}
The exact integral equations for $A_{ap}^\tau$, $A_{ap}^u$, $A_{ap}^{\pi_b}$ and $A_{ap}^{q_b}$ are obtained based on~\eqref{rewritten_KE}, \eqref{f1_structure} and \eqref{mu1_time_equations}:
\begin{eqnarray} \label{mu1_integral_equations}
&& {\lambda _T}A_{ap}^\tau  = \sum\limits_b {{{\hat K}_{ab}}A_{bp}^\tau }\,, \qquad  2{h_{nlp}}\sum\limits_b {{\lambda _{{\pi _a}{\pi _b}}}A_{ap}^{{\pi _b}}}=\sum\limits_b {{{\hat K}_{ab}}{h_{nlp}}A_{bp}^{{\pi _c}}}\,, \nonumber \\
&&A_{ap}^u{p_l}{\lambda _{uu}} + \sum\limits_b {A_{ap}^{{q_b}}{p_l}{\lambda _{{q_b}u}}}  = \sum\limits_b {{{\hat K}_{ab}}{p_l}A_{bp}^u}\,, \nonumber \\
&& A_{ap}^u{p_l}{\lambda _{u{q_c}}} + \sum\limits_b {{p_l}A_{ap}^{{q_b}}{\lambda _{{q_b}{q_c}}}}  = \sum\limits_b {{{\hat K}_{ab}}{p_l}A_{bp}^{{q_c}}} .
\end{eqnarray}
There are additional conditions to the integral equations in~\eqref{mu1_integral_equations}:
\begin{eqnarray} \label{additional_conditions}
&&\langle A_{ap}^\tau\rangle_a  = 0, \qquad \frac{3}{2}{n_a}{s_a} = \langle\varepsilon _{ap}A_{ap}^\tau\rangle_a\,, \qquad   {\delta _{ab}}=\frac{8}{{15}}m_a\langle\varepsilon _{ap}^2  A_{ap}^{{\pi _b}}\rangle_a\,,   \nonumber \\
&&{r_a}{n_a}T = \frac{4}{{15}}\langle{\varepsilon _{ap}^2}A_{ap}^u\rangle_a\,, \qquad {\delta _{ab}} = \frac{2}{{3}}\langle \varepsilon _{ap}^2 A_{ap}^{{q_b}}\rangle_a
\end{eqnarray}
following from the definitions of the RDPs (here, for an arbitrary function $h_p$, the notation $\langle h_p\rangle_a\equiv \int \rd^3 p\,w_{ap}h_p$ is used).

Equations~\eqref{mu1_integral_equations} show that the functions $A_{ap}^\tau$, $A_{ap}^{\pi_b}$ are not coupled with any other functions from these equations. The equation for $A_{ap}^\tau$ completely coincides with the equations for the temperature part of $f_{ap}^{\,(1)}$ obtained in \cite{GorSokCMP}. The equations for $A_{ap}^{q_b}$ and $A_{ap}^u$ are coupled with each other.

The functions $A_{ap}^\tau$,   $A_{ap}^{\pi_b}$, $A_{ap}^{q_b}$ and $A_{ap}^u$ are sought for in a $\sigma$ perturbation theory and expansion in the Sonine polynomials:
\begin{eqnarray} \label{f1_sigma}
&& A_{ap}^\tau  = \sum\limits_{n,s \geqslant 0} {g_{as}^{\tau [n]}S_s^{{1 \mathord{\left/
 {\vphantom {1 2}} \right.
 \kern-\nulldelimiterspace} 2}}\left( {\beta {\varepsilon _{ap}}} \right)},  \qquad  A_{ap}^u = \sum\limits_{n,s \geqslant 0} {g_{as}^{u[n]}S_s^{{3 \mathord{\left/
 {\vphantom {3 2}} \right.
 \kern-\nulldelimiterspace} 2}}\left( {\beta {\varepsilon _{ap}}} \right)}
, \nonumber \\
&& A_{ap}^{{q_b}} = \sum\limits_{n,s \geqslant 0} {g_{as}^{{q_b}[n]}S_s^{{3 \mathord{\left/
 {\vphantom {3 2}} \right.
 \kern-\nulldelimiterspace} 2}}\left( {\beta {\varepsilon _{ap}}} \right)}, \qquad A_{ap}^{{\pi _b}} = \sum\limits_{n,s \geqslant 0} {g_{as}^{{\pi _b}[n]}S_s^{{5 \mathord{\left/
 {\vphantom {5 2}} \right.
 \kern-\nulldelimiterspace} 2}}\left( {\beta {\varepsilon _{ap}}} \right)}
\end{eqnarray}
(here and in what follows, the superscript in brackets denotes the order in $\sigma$). The orthogonal Sonine polynomials are defined by the formula:
\begin{equation} \label{Sonine_polynomials}
S_n^\alpha \left( x \right) \equiv \frac{1}{{n!}}{\re^x}{x^{ - \alpha }}\frac{{{\rd^n}}}{{\rd{x^n}}}\left( {{\re^{ - x}}{x^{\alpha  + n}}} \right)
\end{equation}
and satisfy the orthogonality condition
\begin{equation} \label{Sonine_orthogonality}
\int_0^\infty  {\rd x{\re^{ - x}}{x^\alpha }S_n^\alpha \left( x \right)S_{n'}^\alpha \left( x \right)}  = \frac{1}{{n!}}\Gamma \left( {n + \alpha  + 1} \right){\delta _{nn'}}.
\end{equation}
Such a choice of polynomials in~\eqref{f1_sigma} takes into account the fact that the functions $S^\alpha_s(\beta\varepsilon_{ap})$ are orthogonal with the weight $\varepsilon_{ap}^\alpha w_{ap}$ and the first few coefficients of polynomial expansions~\eqref{f1_sigma} can be obtained from the additional conditions  \eqref{additional_conditions}:
\begin{eqnarray} \label{coefficients_from_additional}
&&g_{e0}^{u\left[ n \right]} = \beta\delta_{n,0}\,,  \qquad  g_{i0}^{u[n]} =  -z\sigma^2\beta\delta_{n,2}\,, \nonumber \\
&& g_{a0}^{{q_b}[n]} = 0, \qquad g_{a1}^{u\left[ {n} \right]} = 0, \qquad g_{a1}^{{q_b}\left[n \right]} =  - \frac{2\beta^2}{{5{n_a}}}{\delta _{ab}}\delta_{n,0}\,, \nonumber \\
&& g_{e0}^{{\pi _b}\left[n \right]} = \frac{\beta^2}{{2{n_e}{m_e}}}\delta _{eb}\delta_{n,0}\,,\qquad g_{i0}^{{\pi _b}\left[n \right]} = \frac{{{\sigma^2\beta^2}}}{{2{n_i}{m_e}}}\delta _{ib}\delta_{n,2}\,,\nonumber\\
&&g_{a0}^{\tau \left[ n \right]} =0, \qquad g_{e1}^{\tau \left[n \right]} =  -\beta\delta_{n,0}\,, \qquad g_{i1}^{\tau \left[n \right]} = z\beta\delta_{n,0}.
\end{eqnarray}
It is easy to see that a standard result~\eqref{standardCDF} is completely given by expressions~\eqref{coefficients_from_additional}. Other contributions to \eqref{f1_sigma} give our corrections to \eqref{standardCDF} in the form of a series in $\sigma$. Therefore, the relaxation constants~\eqref{mu1_relaxation_constants} are also expanded into a series in $\sigma$.

In what follows, the CDFs~\eqref{f1_structure} and coefficients in the time evolution equations for the RDPs~\eqref{mu1_time_equations} are sought for in a $\sigma$ perturbation theory based on \eqref{mu1_integral_equations}, \eqref{f1_sigma}, \eqref{coefficients_from_additional}. This procedure is described in detail in \cite{GorSokCMP} (moreover, the equation for $A_{ap}^\tau$ was solved in \cite{GorSokCMP}). Then, for simplicity, in each order in $\sigma$, the results for the CDFs are found in the one-polynomial approximation, which yields:
\begin{eqnarray} \label{f1_results}
&& A_{ep}^u = \beta + g_{e2}^{u\left[ 2 \right]}S_2^{{3 \mathord{\left/
{\vphantom {3 2}} \right.
\kern-\nulldelimiterspace} 2}}\left( {\beta {\varepsilon _{ep}}} \right) + O\left( {{\sigma ^4}} \right),  \qquad A_{ip}^u =  - z\beta {\sigma ^2} + O\left( {{\sigma ^6}} \right),
\nonumber \\
&& A_{ep}^{{q_e}} =  - \frac{{2\beta^2}}{{5{n_e}}}S_1^{{3 \mathord{\left/
{\vphantom {3 2}} \right.
\kern-\nulldelimiterspace} 2}}\left( {\beta {\varepsilon _{ep}}} \right) + g_{e2}^{{q_e}\left[ 2 \right]}S_2^{{3 \mathord{\left/
 {\vphantom {3 2}} \right.
 \kern-\nulldelimiterspace} 2}}\left( {\beta {\varepsilon _{ep}}} \right) + O\left( {{\sigma ^4}} \right), \qquad A_{ip}^{{q_e}} = O\left( {{\sigma ^6}} \right),
\nonumber \\
&& A_{ip}^{{q_i}} =  - \frac{{2\beta^2}}{{5{n_i}}}S_1^{{3 \mathord{\left/
 {\vphantom {3 2}} \right.
 \kern-\nulldelimiterspace} 2}}\left( {\beta {\varepsilon _{ip}}} \right) + g_{i2}^{{q_i}\left[ 3 \right]}S_2^{{3 \mathord{\left/
 {\vphantom {3 2}} \right.
 \kern-\nulldelimiterspace} 2}}\left( {\beta {\varepsilon _{ip}}} \right) + O\left( {{\sigma ^4}} \right),
\nonumber \\
&& A_{ep}^{{q_i}} = g_{e2}^{{q_i}\left[ 2 \right]}S_2^{{3 \mathord{\left/
 {\vphantom {3 2}} \right.
 \kern-\nulldelimiterspace} 2}}\left( {\beta {\varepsilon _{ep}}} \right) + g_{e2}^{{q_i}\left[ 3 \right]}S_2^{{3 \mathord{\left/
 {\vphantom {3 2}} \right.
 \kern-\nulldelimiterspace} 2}}\left( {\beta {\varepsilon _{ep}}} \right) + O\left( {{\sigma ^4}} \right),
\nonumber \\
&& A_{ep}^{{\pi _e}} = \frac{{\beta^2}}{{2{n_e}{m_e}}} + g_{e1}^{{\pi _e}\left[ 2 \right]}S_1^{{5 \mathord{\left/
 {\vphantom {5 2}} \right.
 \kern-\nulldelimiterspace} 2}}\left( {\beta {\varepsilon _{ep}}} \right) + O\left( {{\sigma ^4}} \right),
\nonumber \\
&& A_{ip}^{{\pi _i}} = \frac{{\beta^2\sigma^2}}{{2{n_i}{m_e}}} + g_{i1}^{{\pi _i}\left[ 5 \right]}S_1^{{5 \mathord{\left/
 {\vphantom {5 2}} \right.
 \kern-\nulldelimiterspace} 2}}\left( {\beta {\varepsilon _{ip}}} \right) + O\left( {{\sigma ^6}} \right), \nonumber \\
&& A_{ep}^{{\pi _i}} = g_{e1}^{{\pi _i}\left[ 2 \right]}S_1^{{5 \mathord{\left/
 {\vphantom {5 2}} \right.
 \kern-\nulldelimiterspace} 2}}\left( {\beta {\varepsilon _{ep}}} \right) + g_{e1}^{{\pi _i}\left[ 3 \right]}S_1^{{5 \mathord{\left/
 {\vphantom {5 2}} \right.
 \kern-\nulldelimiterspace} 2}}\left( {\beta {\varepsilon _{ep}}} \right) + O\left( {{\sigma ^4}} \right), \nonumber \\
&& A_{ip}^{{\pi _e}} = g_{i1}^{{\pi _e}\left[ 6 \right]}S_1^{{5 \mathord{\left/
 {\vphantom {5 2}} \right.
 \kern-\nulldelimiterspace} 2}}\left( {\beta {\varepsilon _{ip}}} \right) + O\left( {{\sigma ^7}} \right), \nonumber \\
&& A_{ep}^\tau  =  - \beta S_1^{{1 \mathord{\left/
 {\vphantom {1 2}} \right.
 \kern-\nulldelimiterspace} 2}}\left( {\beta {\varepsilon _{ep}}} \right) + 3\sqrt 2 z\left( {z + 1} \right)\beta S_2^{{1 \mathord{\left/
 {\vphantom {1 2}} \right.
 \kern-\nulldelimiterspace} 2}}\left( {\beta {\varepsilon _{ep}}} \right){\sigma ^2} + O\left( {{\sigma ^4}} \right),  \nonumber\\
&& A_{ip}^\tau  = z\beta S_1^{{1 \mathord{\left/
 {\vphantom {1 2}} \right.
 \kern-\nulldelimiterspace} 2}}\left( {\beta {\varepsilon _{ip}}} \right) + 2\sqrt 2 \left( {1 + {z^{ - 1}}} \right)\beta S_2^{{1 \mathord{\left/
 {\vphantom {1 2}} \right.
 \kern-\nulldelimiterspace} 2}}\left( {\beta {\varepsilon _{ip}}} \right){\sigma ^3} + O\left( {{\sigma ^4}} \right),
\end{eqnarray}
where the estimates of the type $O\left(\sigma ^n \right)$ are the priori ones.
The numerical data for the coefficients in \eqref{f1_results} are given in table~\ref{table_1} for $z=1,2,3,4$.

\begin{table}[!h]
\caption{Numerical data for the coefficients in \eqref{f1_results}.}
\begin{center}
\begin{tabular}{|c|c|c|c|c|c|} \hline\hline
$z$ & $\frac{{{n_i}}}{{{\beta ^2}{\sigma ^2}}}g_{e2}^{{q_e}\left[ 2 \right]}$ & $\frac{{{n_i}}}{{{\beta ^2}{\sigma ^3}}}g_{i2}^{{q_i}\left[ 3 \right]}$  & $\frac{{{n_i}}}{{{\beta ^2}{\sigma ^2}}}g_{e2}^{{q_i}\left[ 2 \right]}$ & $\frac{{{n_i}}}{{{\beta ^2}{\sigma ^3}}}g_{e2}^{{q_i}\left[ 3 \right]}$ & $\frac{{{n_i}{m_e}}}{{{\beta ^2}{\sigma ^6}}}g_{i1}^{{\pi _e}\left[ 6 \right]}$ \\
\hline\hline
1 & 39 & 4.2 & $-$0.36 & $-$0.082&$-$0.17\\
2 & 0.10 & 2.1 & $-$0.44 & $-$0.49&$-$0.21\\
3 & 0.33 & 1.4 & $-$0.47 & $-$1.3&$-$0.23\\
4 & 0.10 & 1.0 & $-$0.49 & $-$2.5&$-$0.24\\
\hline\hline
$z$ & $\frac{{{n_i}{m_e}}}{{{\beta ^2}{\sigma ^2}}}g_{e1}^{{\pi _e}\left[ 2 \right]}$ & $\frac{{{n_i}{m_e}}}{{{\beta ^2}{\sigma ^5}}}g_{i1}^{{\pi _i}\left[ 5 \right]}$  & $\frac{{{n_i}{m_e}}}{{{\beta ^2}{\sigma ^2}}}g_{e1}^{{\pi _i}\left[ 2 \right]}$ & $\frac{{{n_i}{m_e}}}{{{\beta ^2}{\sigma ^3}}}g_{e1}^{{\pi _i}\left[ 3 \right]}$ & $\frac{1}{{\beta {\sigma ^2}}}g_{e2}^{u\left[ 2 \right]}$ \\
\hline\hline
1 & 0.097& $-$3.2 & 0.14 & 0.047&$-$39\\
2 & 0.061 & $-$1.6 & 0.17 & 0.30&$-$2.8\\
3 & 0.045 & $-$1.1 & 0.19 & 0.81&$-$2.0\\
4 & 0.035 & $-$0.80 & 0.20 & 1.6&$-$1.6\\
\hline\hline
\end{tabular}
\end{center}
\label{table_1}
\end{table}

The leading-in-$\sigma$ terms for $A_{ap}^u$, $A_{ep}^{q_e}$, $A_{ip}^{q_i}$, $A_{ep}^{\pi_e}$, $A_{ip}^{\pi_i}$, $A_{ap}^\tau$ coincide with the standard result \eqref{standardCDF}, but our results~\eqref{f1_results} give corrections in higher orders in $\sigma$ to \eqref{standardCDF}. Although
\begin{equation} \label{ab_lessThan_aa}
A_{ep}^{{\pi _i}} \ll A_{ep}^{{\pi _e}}\,, \qquad A_{ip}^{{\pi _e}} \ll A_{ip}^{{\pi _i}}\,, \qquad A_{ep}^{{q_i}} \ll A_{ep}^{{q_e}}\,, \qquad A_{ip}^{{q_e}} \ll A_{ip}^{{q_i}}\,,
\end{equation}
$A_{ep}^{{\pi _i}}$, $A_{ip}^{{\pi _e}}$, $A_{ep}^{{q_i}}$, $A_{ip}^{{q_e}}$ are not equal to zero in contrast to \eqref{standardCDF}, and the electron distribution function depends on the ion energy and the momentum fluxes and vice versa.

Based on \eqref{mu1_time_equations}, \eqref{mu1_relaxation_constants} and \eqref{f1_results}, the following results for the RDP time evolution equations in the leading and next-to-leading orders in $\sigma$ are obtained:
\begin{eqnarray} \label{1_time_equations}
&&{\left( {{\partial _t}{u_n}} \right)^{(1)}} =  - \sum\limits_{s = 0,2} {\lambda _{uu}^{\left[ s \right]}} {u_n} - \sum\limits_{s = 0,2} {\lambda _{u{q_e}}^{\left[ s \right]}} q_{en}^o - \lambda _{u{q_i}}^{\left[ 2 \right]}q_{in}^o\,,\nonumber \\
&&{\left(  {\partial _t}q_{en}^o \right)^{(1)}} =  - \sum\limits_{s = 0,2} {\lambda _{{q_e}u}^{\left[ s \right]}} {u_n} - \sum\limits_{s = 0,2} {\lambda _{{q_e}{q_e}}^{\left[ s \right]}} q_{en}^o - \lambda _{{q_e}{q_i}}^{\left[ 2 \right]}q_{in}^o\,,\nonumber \\
&&{\left( {{\partial _t}q_{in}^o} \right)^{(1)}} =  - \sum\limits_{s = 1,2} {\lambda _{{q_i}{q_i}}^{\left[ s \right]}} q_{in}^o\,, \qquad {\left( {{\partial _t}\tau } \right)^{(1)}} =  - \sum\limits_{s = 2,4} {\lambda _T^{\left[ s \right]}} \tau,\nonumber \\
&&{\left( {{\partial _t}\pi _{enl}^o} \right)^{(1)}} =  - 2\sum\limits_{s = 0,2} {\lambda _{{\pi _e}{\pi _e}}^{[2]}\pi _{enl}^o}  - 2\lambda _{{\pi _e}{\pi _i}}^{[2]}\pi _{inl}^o\,,\nonumber \\
&&{\left( {{\partial _t}\pi _{inl}^o} \right)^{(1)}} =  - 2\lambda _{{\pi _i}{\pi _e}}^{[2]}\pi _{enl}^o - 2\sum\limits_{s = 1,2} {\lambda _{{\pi _i}{\pi _i}}^{[2]}\pi _{inl}^o}\,,
\end{eqnarray}
where
\begin{eqnarray}  \label{1_relaxation_constants}
&&\lambda _{uu}^{\left[ 0 \right]} = \frac{{4\sqrt 2 {z^2}}}{{3T}}\Lambda, \qquad \lambda _{uu}^{\left[ 2 \right]} = \left[ {\frac{{\left( {4z - 2} \right){\sigma ^2}}}{{3T}} + \frac{5}{2}g_{e2}^{u\left[ 2 \right]}} \right]\sqrt 2 {z^2}\Lambda,
\nonumber \\
&&  \lambda _{u{q_e}}^{\left[ 0 \right]} =  - \frac{{4\sqrt 2 {z^2}}}{{5{n_e}{T^2}}}\Lambda , \qquad \lambda _{u{q_e}}^{\left[ 2 \right]} = \left( {\frac{6{\sigma ^2}}{{5{n_e}{T^2}}} + \frac{5}{2}g_{e2}^{{q_e}\left[ 2 \right]}} \right)\sqrt 2 {z^2}\Lambda,
\nonumber \\
&& \lambda _{u{q_i}}^{\left[ 2 \right]} = \left( {\frac{4{\sigma ^2}}{{5{n_i}{T^2}}} + \frac{5}{2}g_{e2}^{{q_i}\left[ 2 \right]}} \right)\sqrt 2 {z^2}\Lambda, \qquad \lambda _{{q_e}u}^{\left[ 0 \right]} =  - 2\sqrt 2 {n_e}{z^2}\Lambda,
\nonumber \\
&& \lambda _{{q_e}{q_e}}^{\left[ 0 \right]} = \frac{{2\left( {8 + 13\sqrt 2 {z}} \right)}}{{15T}}z\Lambda, \qquad \lambda _{{q_i}{q_i}}^{\left[ 1 \right]} = \frac{{16{z^4}}}{{15T}}\sigma \Lambda, \qquad
\lambda _{{q_i}{q_i}}^{\left[ 2 \right]} = \frac{{4\sqrt 2 {z^3}}}{T}{\sigma ^2}\Lambda,
\nonumber \\
&& \lambda _{{q_e}u}^{[2]} = \left[ {\left( {3 - 2z} \right)\sqrt 2 z{\sigma ^2} - g_{e2}^{u\left[ 2 \right]}T\left( {\frac{{23}}{{2\sqrt 2 }}z + 2} \right)} \right]{z^2}{n_i}\Lambda,
\nonumber \\
&& \lambda _{{q_e}{q_e}}^{\left[ 2 \right]} = \left[ { - \frac{{11\sqrt 2 }}{{5T}}{\sigma ^2} - g_{e2}^{{q_e}\left[ 2 \right]}\left( {\frac{{23}}{{2\sqrt 2 }}z + 2} \right){n_i}T} \right]{z^2}\Lambda, \nonumber\\
&& \lambda _{{q_e}{q_i}}^{\left[ 2 \right]} = \left[ { - \frac{{58\sqrt 2 z}}{{25T}} - \left( {\frac{{23}}{{2\sqrt 2 }}z + 2} \right)g_{e2}^{{q_i}\left[ 2 \right]}{n_i}T} \right]{z^2}\Lambda;
\nonumber\\
&& \lambda _{{\pi _e}{\pi _e}}^{[0]} = \frac{{4z\left( {1 + \sqrt 2 z} \right) }}{{5T}}\Lambda,  \qquad\lambda _{{\pi _i}{\pi _i}}^{[1]} = \frac{{4{z^4} }}{{5T}}\sigma\Lambda, \qquad \lambda _{{\pi _i}{\pi _i}}^{[2]} = \frac{{4\sqrt 2 {z^3} }}{{3T}}{\sigma ^2}\Lambda,
\nonumber \\
&&\lambda _{{\pi _e}{\pi _e}}^{[2]} =\left[ \frac{{2\sqrt 2  {\sigma ^2}}}{{15T}} + \frac{6}{5}\left( {1 + 2\sqrt 2 z} \right){n_i}{m_e}T g_{e1}^{{\pi _e}[2]}\right]z^2\Lambda,
\nonumber \\
&& \lambda _{{\pi _e}{\pi _i}}^{[2]} =  - \left[\frac{{8\sqrt 2 {z} }}{{15T}}{\sigma ^2} + \frac{6}{5}\left( {1 + 2\sqrt 2 z} \right){n_i}{m_e}T g_{e1}^{{\pi _i}[2]}\right]z^2\Lambda,
\nonumber \\
&&\lambda _{{\pi _i}{\pi _e}}^{[2]} =   - \frac{{8\sqrt 2  {\sigma ^2}}}{{15T}}{z^2}\Lambda,\qquad \lambda _T^{\left[ 2 \right]} = \frac{{8\sqrt 2 \sigma^2}}{{3T}}(z+1)z^2\Lambda,
\nonumber \\
&& \lambda _T^{\left[ 4 \right]} =  -\frac{4(\sqrt 2-6z)}{T}\sigma^4(z+1)z^2\Lambda,
\end{eqnarray}
where $\Lambda  \equiv {n_i}{e^4}L (\piup/m_eT)^{1/2}$.

The leading-in-$\sigma$ contributions to the RDP time evolution equations~\eqref{1_time_equations}, \eqref{1_relaxation_constants} are completely defined by the standard CDFs~\eqref{standardCDF}, and in the cases known in the literature they coincide with the known results (see \cite{AlBogRuh,GorSokCMP}).

As for the next-to-leading terms in \eqref{1_time_equations}, \eqref{1_relaxation_constants}, we have the following. The quantities $\lambda_{q_iq_i}^{[2]}$, $\lambda_{\pi_i\pi_i}^{[2]}$, $\lambda_{\pi_i\pi_e}^{[2]}$ are completely defined by a standard result~\eqref{standardCDF}, and the time evolution equations for $q_{in}^o$ and $\pi_{inl}^o$ in the leading and next-to-leading terms in $\sigma$ are completely defined by \eqref{standardCDF}. However, the other time evolution equations are not completely defined by  \eqref{standardCDF}. In the right-hand sides of the expressions for $\lambda_{uu}^{[2]}$, $\lambda_{uq_e}^{[2]}$, $\lambda_{uq_i}^{[2]}$, $\lambda_{q_eu}^{[2]}$, $\lambda_{q_eq_e}^{[2]}$, $\lambda_{\pi_e\pi_e}^{[2]}$, $\lambda_{\pi_e\pi_i}^{[2]}$, $\lambda_{T}^{[4]}$, the first terms come from the standard CDFs~\eqref{standardCDF}, but the second terms come from our corrections~\eqref{f1_results}. A comparison of these terms is given in table~\ref{table_2}. Thus, our corrections have a significant effect on the next-to-leading terms in the RDP time evolution equations. These next-to-leading terms coincide with the known results in the cases known in the literature (see \cite{Ishi,GorSokCMP}).

\begin{table}[!h]
\caption{Comparison of the magnitudes of the terms which come from our corrections $\lambda''$ and the terms which come from \eqref{standardCDF} $\lambda'$.}
\vspace{2ex}
\begin{center}
\begin{tabular}{|c|c|}
\hline\hline
Quantity & Comparison \\
\hline\hline
$\lambda_{T}^{[4]}$ & $\lambda_{T}^{[4]''} > \lambda_{T}^{[4]'}$  \\
$\lambda_{uq_i}^{[2]}$ &$\lambda_{uq_i}^{[2]''}>\lambda_{uq_i}^{[2]'}$\\
$\lambda_{\pi_e\pi_e}^{[2]}$ & $\lambda_{\pi_e\pi_e}^{[2]''}>\lambda_{\pi_e\pi_e}^{[2]'}$\\
$\lambda_{uu}^{[2]}$ & $\lambda_{uu}^{[2]''}>\lambda_{uu}^{[2]'}$ for $z=1,2,3$ \\
$\lambda_{uq_e}^{[2]}$ & $\lambda_{uq_e}^{[2]''}>\lambda_{uq_e}^{[2]'}$ for $z=1,2,3$ \\
$\lambda_{q_eu}^{[2]}$ & $\lambda_{q_eu}^{[2]''}>\lambda_{q_eu}^{[2]'}$ for $z=1,2,3,4,5$ \\
$\lambda_{q_eq_e}^{[2]}$ & $\lambda_{q_eq_e}^{[2]''}>\lambda_{q_eq_e}^{[2]'}$ for $z=1,2,3,4$ \\
$\lambda_{q_eq_i}^{[2]}$ & $\lambda_{q_eq_i}^{[2]''}<\lambda_{q_eq_i}^{[2]'}$ but they are comparable\\
$\lambda_{\pi_e\pi_i}^{[2]}$ & $\lambda_{\pi_e\pi_i}^{[2]''}<\lambda_{\pi_e\pi_i}^{[2]'}$ but they are comparable \\
\hline\hline
\end{tabular}
\end{center}
\label{table_2}
\end{table}

It is interesting to note that numerically for $z=1$ (for example, in the case of electron-proton plasma) our corrections to the time evolution equations are most significant for  ${( {{\partial _t}{u_n}})^{(1)}}$ and ${( {{\partial _t}q_{en}^o})^{(1)}}$. Namely, although for this plasma ${\sigma ^2} = 5.5 \cdot {10^{ - 4}}$ we have
\begin{eqnarray} \label{ratios}
&& \frac{{\lambda _{uu}^{[2]}}}{{\lambda _{uu}^{[0]}}} =  - 4.0 \cdot {10^{ - 2}}, \qquad \frac{{\lambda _{u{q_e}}^{[2]}}}{{\lambda _{u{q_e}}^{[0]}}}=  - 2.2 \cdot {10^{ - 2}},  \nonumber \\
&& \frac{{\lambda _{{q_e}u}^{[2]}}}{{\lambda _{{q_e}u}^{[0]}}} =  - 2.6 \cdot {10^{ - 2}}, \qquad \frac{{\lambda _{{q_e}{q_e}}^{[2]}}}{{\lambda _{{q_e}{q_e}}^{[0]}}}=  - 2.0 \cdot {10^{ - 2}},
\end{eqnarray}
the ratios~\eqref{ratios} are of the order $10^{-2}$, rather than $10^{-4}$.

\section{Conclusions}

This paper is devoted to a generalization of the 13-moment Grad approximation for a spatially uniform completely ionized two-component electron-ion plasma that describes the Maxwell relaxation. The investigation is based on the Landau kinetic equation and our generalization of the Chapman--Enskog method \cite{GorSokNasMath} with the help of the Bogolyubov idea of the functional hypothesis, which is the main idea of his method of a reduced description.

The system is considered to be in the vicinity of its equilibrium state where the deviations of the reduced description parameters from their equilibrium values are small. This introduces a small parameter~$\mu$ which allows us to calculate the component non-equilibrium distribution functions of a system and obtain time evolution equations for the reduced description parameters in a corresponding perturbation theory. In contrast to the standard Grad method, our small parameter $\mu$ allows us to obtain the CDFs based on the kinetic equation.

In the present paper, the investigation is restricted to a relaxation theory linear in $\mu$ for spatially uniform states. However, our approach allows one to investigate a nonlinear relaxation in non-uniform systems \cite{GorSokNasMath,GorSokIJMPB,GorSokVis}.

Our results for the component distribution functions are compared with the standard results of the Grad method in plasma physics \cite{Silin} given by expression~\eqref{standardCDF}. We first calculated the component distribution functions in a perturbation theory in the small square root of the electron-to-ion mass ratio $\sigma$. Then, in each order in $\sigma$, we restricted ourselves to the Sonine one-polynomial approximation for simplicity. It is obtained that the leading-in-$\sigma$ results for the component distribution functions coincide with the standard result~\eqref{standardCDF}, but corrections to it in higher orders in $\sigma$ are obtained. Moreover, it is obtained that in contrast to \eqref{standardCDF}, the electron distribution function depends on the ion energy and momentum fluxes and vice versa, although this dependence takes place in higher-than-leading orders in $\sigma$.

Time evolution equations for the reduced description parameters are also calculated. They are obtained in the leading-in-$\sigma$ and next-to-leading orders. It is obtained that their leading-in-$\sigma$ terms are completely defined by the standard CDFs~\eqref{standardCDF}. The next-to-leading terms in the time evolution equations for the ion fluxes $q_{in}^o$ and $\pi_{inl}^o$ are completely defined by \eqref{standardCDF}, but our corrections to the component distribution functions have a significant effect on the next-to-leading terms in the time evolution equations for the deviations $\tau$, $u_n$ of the temperatures, velocities  and electron fluxes $q_{en}^o$, $\pi_{enl}^o$.

The Grad method is widely used in modern statistical physics and the idea of the paper may be applied to its generalization not only for plasma, but also for other systems. Moreover, the obtained results may be the basis for the investigation of spatially non-uniform states of plasma because they are the results of the leading order in small gradients.

\ukrainianpart

\title{Узагальнення метода Греда в фізиці плазми}
\author{В.М. Горєв, О.Й. Соколовський}
\address{
Дніпровський національний університет імені Олеся Гончара, пр. Гагаріна, 72, 49010 Дніпро, Україна}

\makeukrtitle

\begin{abstract}
\tolerance=3000%
Метод Греда узагальнюється на основі ідеї функціональної гіпотези Боголюбова для станів наприкінці завершення релаксаційних процесів у системі. Для повністю іонізованої однорідної двокомпонентної електрон-іонної плазми за допомогою кінетичного рівняння Ландау досліджується проблема Греда (опис максвеллівської релаксації). Обчислюються функція розподілу компонент і часові рівняння для параметрів, які описують стан системи, знаходяться корекції до відомих результатів в теорії збурень за малим відношенням мас електрона до іона.%
\keywords Максвеллівська релаксація, метод Греда, узагальнений метод Чепмена-Енскога, повністю іонізована плазма, поліноми Соніна
\end{abstract}

\end{document}